\documentclass[aps,prb,twocolumn,superscriptaddress,longbblliography]{revtex4-2}

\usepackage[colorlinks=true,
            citecolor=blue,
            linkcolor=blue,
            urlcolor=blue,
            breaklinks=true]{hyperref}
\usepackage{amssymb}
\usepackage{amsmath}
\usepackage{graphicx}
\usepackage{xcolor}

\usepackage{times} %this package is deprecated... is there a good reason that we are using it? Note if it is removed the figures will need to be moved a bit to get them to be in reasonable places on the pages.

\newcommand{\mh}{\textcolor{black}}

\begin{document}

\title{Evidence for easy-plane XY ferromagnetism in heavy-fermion quantum-critical CeRh$_6$Ge$_4$}

\author{Riku Yamamoto}
\thanks{These authors have equally contributed to this work.}
\affiliation{Materials Physics and Applications $-$ Quantum, Los Alamos National Laboratory, Los Alamos, New Mexico  87545, USA}
%\altaffiliation{Present address: Department of Physics and Astronomy, UCLA, Los Angeles, California 90095, USA. email:ryamamoto@physics.ucla.edu}

\author{Sejun Park}
\thanks{These authors have equally contributed to this work.}
\affiliation{Materials Physics and Applications $-$ Quantum, Los Alamos National Laboratory, Los Alamos, New Mexico 87545, USA}

\author{Zachary W. Riedel}
\affiliation{Materials Physics and Applications $-$ Quantum, Los Alamos National Laboratory, Los Alamos, New Mexico 87545, USA}

\author{Phurba Sherpa}
\affiliation{Materials Physics and Applications $-$ Quantum, Los Alamos National Laboratory, Los Alamos, New Mexico 87545, USA}
\affiliation{Department of Physics and Astronomy, University of California, Davis, California 95616, USA}

\author{Joe D. Thompson}
\affiliation{Materials Physics and Applications $-$ Quantum, Los Alamos National Laboratory, Los Alamos, New Mexico 87545, USA}

\author{Filip Ronning}
\affiliation{Materials Physics and Applications $-$ Quantum, Los Alamos National Laboratory, Los Alamos, New Mexico  87545, USA}

\author{Eric D. Bauer}
\affiliation{Materials Physics and Applications $-$ Quantum, Los Alamos National Laboratory, Los Alamos, New Mexico  87545, USA}

\author{Adam P. Dioguardi}\thanks{apd@lanl.gov}
\affiliation{Materials Physics and Applications $-$ Quantum, Los Alamos National Laboratory, Los Alamos, New Mexico 87545, USA}

\author{Michihiro Hirata}\thanks{mhirata@lanl.gov}
\affiliation{Materials Physics and Applications $-$ Quantum, Los Alamos National Laboratory, Los Alamos, New Mexico  87545, USA}

\date{\today}

\begin{abstract}
We report ${}^{73}$Ge nuclear quadrupole resonance (NQR) and magnetic resonance (NMR) spectroscopy in the heavy-fermion quantum-critical ferromagnet CeRh$_6$Ge$_4$. NQR and NMR spectral measurements at the two non-equivalent Ge sites reveal electric field gradient tensors and the directions of their principal axes relative to the hexagonal basal plane. %that are reasonably %that---
% compared with density functional theory calculations. %---show relatively better agreement with localized 4$f$-electrons.
 The spin-lattice relaxation rate $1/T_1$ experiments reveal a clear critical slowing down approaching the ferromagnetic transition. %accompanied by a second-order phase transition. 
$1/T_1$ in the paramagnetic state is found to be predominantly caused by fluctuating 4$f$-local moments. The Knight shift shows Curie-Weiss behavior at high temperature and a deviation from this below $T^* \approx$ 25\,K possibly due to a mixture of crystalline electric field effects and Kondo screening. % due to either depopulation effects of low-lying crystalline electric field doublets or Kondo hybridization or both. 
Order-parameter-like behavior of hyperfine fields at the Ge sites and ferromagnetic signal enhancement are observed in Zeeman-perturbed NQR, %zero-field NMR, 
with uniform ferromagnetic order and a small ordered moment of $\approx$\,0.26\,$\mu_\mathrm{B}$/Ce confined within the $ab$-plane. The ordered moment shows a notable in-plane magnetic stiffness against out-of-plane radiofrequency fields and has an (XY-type) in-plane isotropic nature. 
Our results reveal a strong easy-plane anisotropy of 4$f$-electron moment and suggest an involved interplay of hybridization and local moment physics in this quantum critical heavy-fermion ferromagnet. 
\end{abstract}
\maketitle

\section{Introduction}

In heavy-fermion systems, an intertwined competition between localization and itinerancy of $f$-electrons brings about a rich variety of physical phenomena, ranging from complicated magnetism and unconventional superconductivity to quantum criticality~\cite{Coleman_2001, Yang_2008, Coleman_2010, Paschen2021}. 
The heavy-fermion ferromagnet CeRh$_6$Ge$_4$, with %a Kondo temperature $T_{\mathrm{K}}\sim19$\,K and 
a Curie temperature $T_\mathrm{C}$ $=$ 2.5\,K at ambient pressure~\cite{Matsuoka}, provides a platform in which such a competition of 4$f$-electrons leads to an unusual ferromagnetic phase transition whose $T_\mathrm{C}$ is tuned to absolute zero under hydrostatic pressure ($P$), giving rise to a $P$-tuned ferromagnetic (FM) quantum critical point (QCP) at $P_\mathrm{C} \sim$  0.8\,GPa~\cite{Shen,Kotegawa,Thomas}. 

Unlike in its antiferromagnetic (AFM) counterparts, stabilizing a FM QCP is extremely difficult because the continuous transition is generally superseded by a discontinuous first-order line~\cite{Belitz_PRL}, resulting in a pair of ``wings'' of phase transition lines that are terminated by quantum critical end points \cite{MFC}. 
FM QCPs are proposed within standard frameworks of order parameter criticality~\cite{Hertz_1976,Millis_1994,Moriya_book} under limited conditions though~\cite{MFC}, including a strong spin-orbit coupling in a noncentrosymmetric lattice~\cite{Kirkpatrick} or being strongly disordered~\cite{Belitz_PRL}. 

There is another framework based on local theory of criticality~\cite{Si_2001} that allows FM QCPs if the charge and spin degrees of freedom have one-dimensional character~\cite{Shen}.
The Kondo-lattice system CeRh$_6$Ge$_4$ is intriguing in this context because its continuous FM transition at ambient pressure is believed to persist up to $P_\mathrm{C}$ without being interrupted by a first-order transition~\cite{Shen}, and the system is supposed to be extremely clean, with a residual resistivity ratio (RRR) ranging from 30 to 50~\cite{Shen,Kotegawa,Thomas}. 
The low-temperature residual resistivity $\rho_0$, the coefficient of the quadratic term in the resistivity $A$, and the heat capacity Sommerfeld coefficient $\gamma$ diverge
upon approaching $P_\mathrm{C}$ and non-Fermi-liquid behavior develops around $P_\mathrm{C}$~\cite{Shen,Thomas}, all of which are suggestive of a FM QCP.
In the work by Shen $\textit{et al.}$~\cite{Shen}, the authors argued that local Kondo-breakdown-type criticality~\cite{Si_2001,Coleman_2001} is realized due to an easy-plane anisotropy of rather localized 4$f$-electron moments forming quasi-one-dimensional (Q1D) chains.
In this framework, easy-plane anisotropy plays an equivalent role as that of magnetic frustration in AFM systems and generates macroscopic entanglement of spins~\cite{Komijani_PRL};
the $c$--$f$ hybridization then enables tuning of the magnetic ordering temperature to $T = 0$\,K at the QCP, where the Fermi surface volume increases from small for $P<P_\mathrm{C}$ to large at $P>P_\mathrm{C}$. 
However, the mechanism driving such a QCP and the understanding about its nature is still controversial
~\cite{Wang,Joe_2021,Thomas}.

At ambient $P$, both localized and itinerant characters of 4$f$ electrons are reported. 
The crystalline-electric-field (CEF) analysis of magnetic susceptibility revealed low-lying CEF doublets with a level separation of 5.8\,meV~\cite{Shu}, while %and 
inelastic neutron scattering found a broad excitation with a full width at half maximum of $\sim$~30\,meV that implies a Kondo scale that is larger than the CEF splitting% mixing of low-lying doublets through $c$--$f$ hybridization
~\cite{Shu}. 
The presence of Kondo hybridization is further suggested by a sizable Sommerfeld coefficient in the ordered state ($\gamma\sim$ 400\,mJ\,mol$^{-1}$\,K$^{-2}$), magnetic entropy recovering little at $T_\mathrm{C}$ (19\,$\%$ of $R\ln2$), and a small ordered moment size (0.2\,--\,0.3\,$\mu_\mathrm{B}$) relative to the expectation from a CEF fit ($\langle \mu_{ab}\rangle$\,$=$\,1.28\,$\mu_\mathrm{B}$)~\cite{Matsuoka,Shen}. 
Note that these features could imply itinerant magnetism or strong quantum fluctuations of local moments from its Q1D structure~\cite{Komijani_PRL}.
Angle-resolved photoemission spectroscopy (ARPES) measurements find an anisotropic $c$--$f$ hybridization% that develops predominantly in the first-excited CEF doublet
~\cite{Wu_ARPEX}.
All these findings point to a rather involved nature of 4$f$ electrons and call for further investigations. 
In particular, understanding of the local magnetic anisotropy and magnetic structure is limited. 

In this work, we report a microscopic investigation of local magnetism in CeRh$_6$Ge$_4$ with nuclear quadrupole resonance (NQR) and magnetic resonance (NMR) spectroscopy. 
Spectral measurements reveal clear evidence for a ferromagnetic order parameter evolution below $T_\mathrm{C}$ %with a notable easy-plane FM anisotropy 
with a small ordered moment of $\approx$\,0.26\,$\mu_\mathrm{B}$/Ce.
The ordered moment is confined within the $ab$-plane, with %resulting in 
a clear magnetic stiffness against out-of-plane radiofrequency (rf) fields and an isotropic (XY-type) in-plane nature.
The temperature independence of the spin-lattice relaxation rate $1/T_1$ in the paramagnetic (PM) state points to local-moment-dominated fluctuations, and the Knight shift $K$ measurements provide hints for a possible interplay of Kondo hybridization and CEF level mixing effects.

\begin{figure}[t]
	\includegraphics[width=8.6cm]{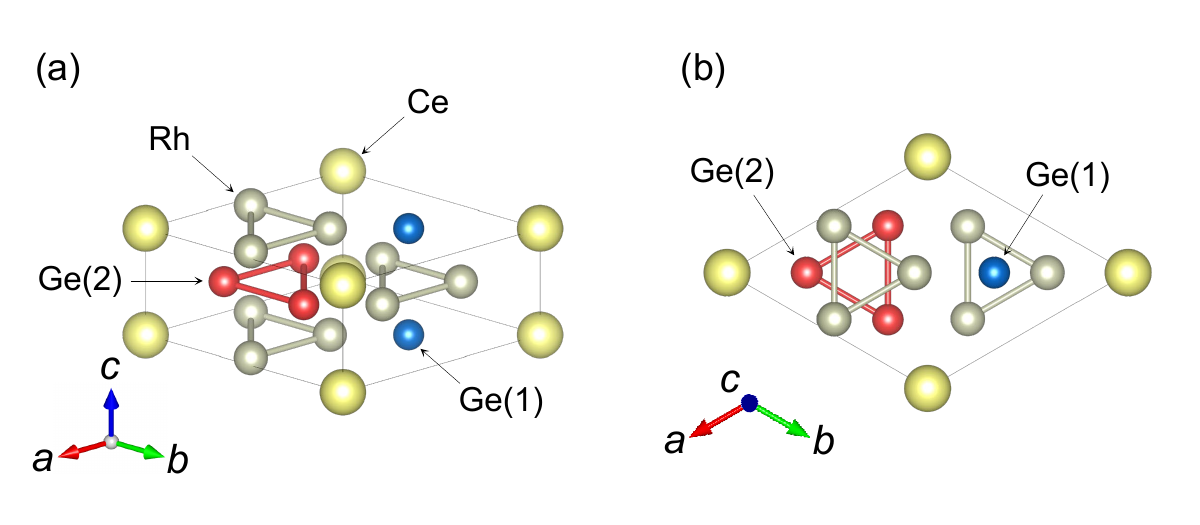}
	\vskip -0.25cm
	\caption{Hexagonal crystal structure of CeRh$_6$Ge$_4$ seen from side (a) and from top (b) views. Yellow balls stand for Ce atoms; gray balls are Rh atoms; and blue and red balls are the two non-equivalent Ge atoms, Ge(1) and Ge(2), respectively. %Also shown are the principal axes of the electric field gradient (EFG) tensor determined by DFT calculations for Ge(1) and Ge(2) given by the green, blue, and red arrows, corresponding to the three principal components of the tensor, $V_{xx}^{(j)}$, $V_{yy}^{(j)}$, and $V_{zz}^{(j)}$, respectively ($V_{zz}^{(j)} \ge V_{yy}^{(j)} \ge V_{xx}^{(j)}$). (b) The same structure seen from the $c$ axis. $V_{zz}^{\mathrm{Ge(1)}}$ is pointing along the $c$ axis, whereas $V_{zz}^{\mathrm{Ge(2)}}$ lies within the $ab$-plane.
    }
    
	\label{fig:intro}
\end{figure}

\section{Sample preparation and characterization}
Single-crystal samples of CeRh$_6$Ge$_4$ were synthesized from Bi flux~\cite{Voßwinke}. The samples were enriched with the NMR/NQR-active ${}^{73}$Ge isotope (nuclear spin $I$\,$=$\,9/2) to 95.60\,$\%$, whereas the natural abundance is 7.73\,$\%$~\cite{Harris}. Synthesis employing ${}^{73}$Ge did not affect the ferromagnetic transition temperature $T_\mathrm{C}$ ($=$\,2.51\,K) relative to that in as-grown samples~\cite{Kotegawa,Shen}. To facilitate penetration of an
rf field, powdered crystals were aligned in a magnetic field to produce a pseudo-single crystal for NMR measurements, while they were randomly oriented for NQR measurements [see the Supplemental material (SM)~\cite{SM} (\mh{see also references \cite{Vannette2008,Hardy1993,CP,MG,Farrell,Takigawa,Curro_LSCO_2000,SmallTip_T1,Chepin_1991,Bordelon2025} for details}).

NQR and NMR measurements were performed at two $^{73}$Ge sites (nuclear gyromagnetic ratio $^{73}\gamma_n/2\pi$\,$=$\,1.4897\,MHz/T and quadrupole moment $^{73}Q$\,$=$\,$-0.196$\,barn~\cite{Harris}) using standard spin-echo and Carr-Purcell-Meiboom-Gill (CPMG) pulse sequence techniques~\cite{CP,MG} with commercially available spectrometers (REDSTONE, Tecmag Inc.) and conventional helium-4 superconducting magnets.  
Measurements above 1.8\,K were conducted using a variable temperature insert cryostat, while below 1.5\,K we used a dilution %fridge 
refrigerator. 
Below 4.0\,K, the samples were immersed in the liquid helium %to have good thermal contact with cryogen 
to minimize the heating.

The two Ge nuclei in CeRh$_6$Ge$_4$ reside at non-cubic lattice positions, one at an axially-symmetric site [which we call Ge(1)] and the other at a non-symmetric site [Ge(2)], as illustrated in Fig.~\ref{fig:intro}.  
Both sites are subjected to a non-vanishing electric field gradient (EFG) that couples to their nuclear quadrupole moment $Q$ %(nonzero for $I \geq 1 $) 
and generates an electric quadrupole interaction. Combined with the nuclear Zeeman term, the %total nuclear spin 
Hamiltonian for the nuclear spin site $j$  is given by:
\begin{equation}
  \begin{aligned}
    \mathcal{H}_{N}^{(j)} & = \mathcal{H}_{Q}^{(j)} + \mathcal{H}_{Z}^{(j)} \\
                                                        &  =  \frac{h\nu_{\mathrm{Q}}^{(j)}}{6}[3I_z^2-I^2+\eta^{(j)}       (I_x^2-I_y^2)]+\gamma_n\hbar \textbf{I}\cdot \textbf{H}^{(j)},
     \label{eq_NQR_Zeeman}
   \end{aligned}
\end{equation} 
with the first and second terms corresponding to the quadrupole and Zeeman terms, respectively. Here, \mh{we define $I^2  = I_x^2 + I_y^2 + I_z^2$,} $\nu_{\mathrm{Q}}^{(j)} \equiv 3eV_{zz}^{(j)}Q/2I(2I-1)h$ is the nuclear quadrupole frequency with the Planck constant $h$ and the elementary charge $e$. $V_{zz}^{(j)}$ is the largest principal component of the EFG tensor, \mh{$\eta^{(j)}=(|V_{yy}^{(j)}|-|V_{xx}^{(j)}|)/|V_{zz}^{(j)}|$} is the asymmetry parameter \mh{($|V_{zz}^{(j)}| \ge |V_{yy}^{(j)}| \ge |V_{xx}^{(j)}|$)}, $\gamma_n$ is the nuclear gyromagnetic ratio, and $\textbf{H}^{(j)}$ is the local magnetic field vector with a magnitude of $H^{(j)}=|\textbf{H}^{(j)}|$. Spectra were analyzed and fit by %making 
exact diagonalization of Eq.~(\ref{eq_NQR_Zeeman}).

%Density functional theory (DFT) calculations were conducted with the generalized gradient approximation and the Perdew-Burke-Ernzerhof exchange correlation functional to estimate the sizes of EFG at two Ge sites. We calculated two cases; (i) Ce 4$f$-electrons are assumed to be fully localized and (ii) are assumed to be fully itinerant.

\section{Results and discussions}

\begin{figure}[t]
	\includegraphics[width=8.6cm]{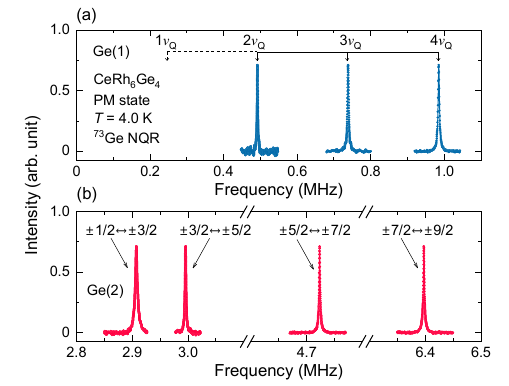}
	\vskip -0.25cm
	\caption{Zero-field $^{73}$Ge NQR spectra of CeRh$_6$Ge$_4$ at 4.0 K in the PM state. The spectrum for (a) Ge(1) and (b) Ge(2) are depicted. The lowest-frequency (1$\nu_{\mathrm{Q}}$) line for Ge(1) (expected at 0.246~MHz) is missing due to the frequency limitation of our spectrometer.}   
	\label{fig:NQR}
\end{figure}

Figures~\ref{fig:NQR}(a) and \ref{fig:NQR}(b) show %typical 
zero-field $^{73}$Ge NQR spectra of CeRh$_6$Ge$_4$ recorded in the PM state at $T =$ 4.0\,K. Two groups of resonance lines were observed in a frequency range between 0.5 and 6.4\,MHz from the two non-equivalent Ge nuclei [Ge(1) and Ge(2) as given in Fig.~\ref{fig:intro}].
The first group of three equally spaced lines appear below 1\,MHz [Fig.~\ref{fig:NQR}(a)].
Considering the local-site symmetries at Ge sites, these three lines are assigned to $I_z =\pm3/2 \leftrightarrow \pm5/2$ (2$\nu_{\mathrm{Q}}$), $\pm5/2 \leftrightarrow \pm7/2$ (3$\nu_{\mathrm{Q}}$), and $\pm7/2 \leftrightarrow \pm9/2$ (4$\nu_{\mathrm{Q}}$) transitions of Ge(1), characterized by EFG parameters of $\nu_{\mathrm{Q}}^\mathrm{Ge(1)}$\,$=$\,0.2463\,MHz and $\eta^\mathrm{Ge(1)}$\,$=$\,0 (i.e., axially symmetric). 
The lowest-frequency 1$\nu_{\mathrm{Q}}$ line is missing because of the frequency limitation of our spectrometer. 
(Note that these frequencies lie near the lowest observable limits for NQR measurements in correlated electron systems and, indeed, to the best of our knowledge 2$\nu_{\mathrm{Q}}^\mathrm{Ge(1)}$\,$=$\, 0.492\,MHz is the lowest frequency reported~\cite{Mito_CeB6}.) 
Between 2.8 and 6.5\,MHz, another set of non-equally spaced four lines were observed and assigned to the four transitions of Ge(2) [Fig.~\ref{fig:NQR}(b)]. By numerically diagonalizing Eq.~(\ref{eq_NQR_Zeeman}) for zero field and fitting to the four lines, we obtained $\nu_{\mathrm{Q}}^\mathrm{Ge(2)} = 1.615$\,MHz and $\eta^\mathrm{Ge(2)} = 0.365$. (Note that these EFG parameters show little temperature dependence below 60\,K; see the SM~\cite{SM}.)

%Table~\ref{tab_NuQ} shows a comparison of the experimentally determined EFG parameters with their calculated values using DFT techniques, assuming either 4$f$-electrons to be fully localized or itinerant. The principal axes of the EFG tensor at Ge(1) and Ge(2) derived from this DFT calculation are also illustrated in Fig.\,\ref{fig:intro}(a). \mhcorr{The calculated $\nu_{\mathrm{Q}}$ values give a fairly good order of estimates for either cases.} 

%\begin{table}[t]
%\caption{Comparison of experimental and DFT-calculated values of EFG parameters at Ge(1) and Ge(2) in CeRh$_6$Ge$_4$.}
%\begin{tabular}{c|c|c|c|c}
%\hline
% \multirow{2}{*}{}    & &\ \ Experiment\ \ &\ \ Localized\ \ &\ \ Itinerant\ \ \\
%                      & & (at 4 K) & (DFT) &(DFT) \\
%\hline
%\multirow{2}{*}{\ Ge(1)\ } & $\ \nu_Q\ $ (MHz)& 0.246&0.276&0.138\\
%                   & $\eta$ &0 &0&0\\
%\hline
%\multirow{2}{*}{\ Ge(2)\ } & $\ \nu_Q$ (MHz)\ &  1.615&1.481& 1.441 \\
%                   & $\eta$&0.365&0.468&0.416  \\
%\hline
%\end{tabular}
%\label{tab_NuQ}
%\end{table}

\label{NMR}
\begin{figure}[t]
	\includegraphics[width=8.0cm]{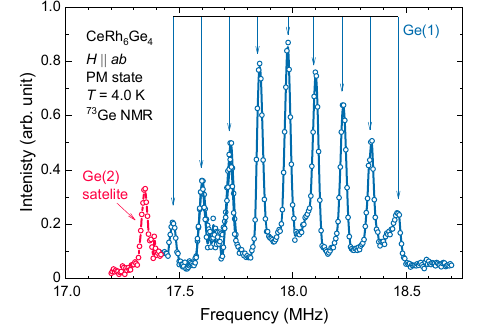}
	\caption{$^{73}$Ge NMR spectrum of CeRh$_6$Ge$_4$ recorded at 4.0\,K under 12\,T using powdered samples that are self-aligned to $H$\,$||$\,$ab$ due to the easy-plane anisotropy. Nine peaks from the axially-symmetric Ge(1) site (blue) is visible, with evenly spaced lines separated by 0.123\,MHz.
The overall background signal and the lowest-frequency peak (red) originate from the Ge(2) site (see the SM for details~\cite{SM}).}
	\label{fig:NMR_full_spectrum}
\end{figure}

\begin{figure}[t]
	\includegraphics[width=8.0cm]{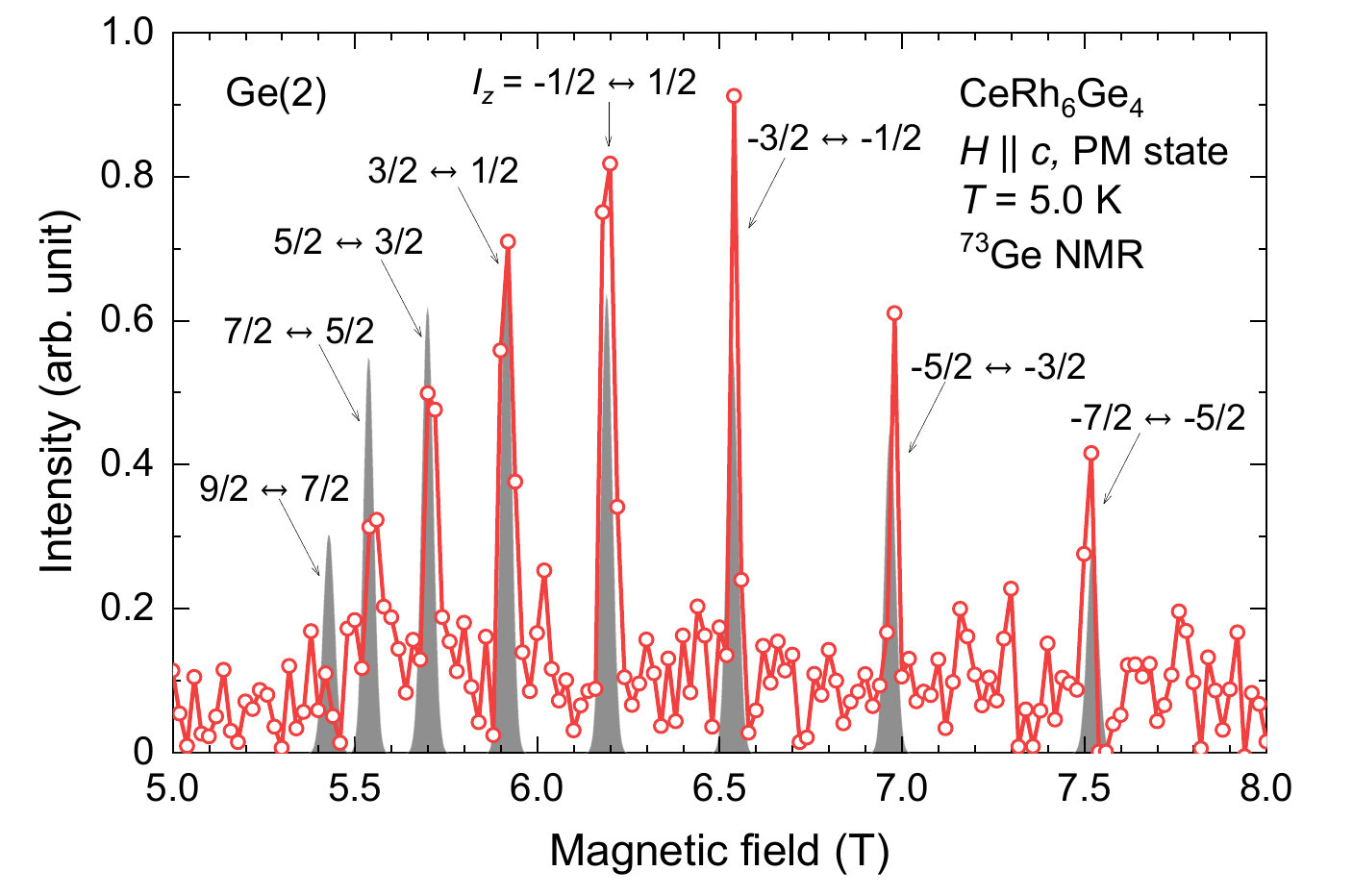}
	\caption{$^{73}$Ge NMR spectrum of CeRh$_6$Ge$_4$ (red) using single crystal samples that are aligned along the $c$-axis, recorded at 9.70\,MHz and 5.0\,K for $H$\,$||$\,$c$. Shaded pattern (gray) gives the calculated single-crystalline spectrum of Ge(2) by exact diagonalization of Eq.~(\ref{eq_NQR_Zeeman}), assuming $\nu_{\mathrm{Q}}^\mathrm{Ge(2)} = 1.615$\,MHz, $\eta^\mathrm{Ge(2)} = 0.365$, $V_{zz}^\mathrm{Ge(2)}$\,$\parallel$\,$ab$, $V_{yy}^\mathrm{Ge(2)}$\,$\parallel$\,$ab$, and $V_{xx}^\mathrm{Ge(2)}$\,$\parallel$\,$c$. The $I_z = -9/2  \leftrightarrow -7/2$ and $9/2  \leftrightarrow 7/2$ lines as well as the Ge(1) signal are missing due to the low signal-to-noise ratio and field range (see the SM for details~\cite{SM}).}
	\label{fig:Ge2-NMR-out-of-plane-field}
\end{figure}

In Fig.~\ref{fig:NMR_full_spectrum} we show a representative $^{73}$Ge NMR spectrum recorded in the PM state at $T$\,$=$\,4.0\,K and $H$\,$=$\,12\,T, using powdered samples that are self-\,aligned to $H$\,$\parallel$\,$ab$ due to the easy-plane anisotropy ($\chi_{ab}/\chi_c \sim 10$ at 4.0\,K~\cite{Shen}, see the SM~\cite{SM} for detail). 
Nine single-crystal-like sharp peaks with a symmetric pattern around the center and an even spacing of  0.123\,MHz is clearly visible. 
Within a first-order perturbative framework of the quadrupole effect for zero asymmetry ($\eta = 0$), the nearest-neighbor NMR line separation in a single crystal is given by $\Delta \nu = \nu_{\mathrm{Q}}(3 ~\mathrm{cos}^2 \theta - 1)/2$, where $\theta$ stands for the angle between the $V_{zz}$ direction and $H$~\cite{Abragam1961}. 
The symmetric pattern of the observed nine-peak spectrum suggests that these lines can be assigned to quadrupole-split, pseudo-single-crystalline NMR peaks coming from the axially-symmetric Ge(1), associated with eight satellite transitions and one central line. 
By comparing with the NQR spectrum in Fig.~\ref{fig:NQR}(a), the line separation yields $\Delta \nu$\,$\approx$\,$\nu_{\mathrm{Q}}^\mathrm{Ge(1)}/2$, which points to $\theta = 90^\circ$; thus, the NMR data unambiguously demonstrate $V_{zz}^{\mathrm{Ge(1)}}$\,$||$\,$c$ ($\perp$\,$H$). %, in excellent agreement with the DFT calculation [see Fig.\,\ref{fig:intro}(a)].  
The last point is also consistent with our observation that there is no second-order quadrupole corrections to %in 
the central line, as anticipated for $\theta = 90^{\circ}$~\cite{Abragam1961}.
(The small underlying asymmetry seen in the peaks may reflect a small misalignment of the crystallites.) 
Outside the frequency range of these nine peaks, we additionally found a complicated spectral pattern and a broad background signal (see the SM~\cite{SM}). 
Such complex features are attributed %should be tied 
to the powder pattern of the Ge(2) signal reflecting the non-axial site symmetry ($\eta^{\mathrm{Ge(2)}} \ne 0$) and the $V_{zz}^{\mathrm{Ge(2)}}$ axis lying off the crystalline $c$-axis direction. 
%[The breadth of the Ge(2) spectrum exceeds the field range we can cover in our magnet, and therefore a full spectral analysis is difficult.] 
In fact, through additional NMR experiments for $H$\,$\parallel$\,$c$ using a mosaic of $c$-axis-aligned single crystals and exact diagonalization of Eq.~(\ref{eq_NQR_Zeeman}), as shown in Fig.~\ref{fig:Ge2-NMR-out-of-plane-field}, we revealed that $V_{xx}^{\mathrm{Ge(2)}}$\,$\parallel$\,$c$ and $V_{yy}^{\mathrm{Ge(2)}}$ and $V_{zz}^{\mathrm{Ge(2)}}$ lie within the $ab$ plane ($V_{yy}^{\mathrm{Ge(2)}}$and $V_{zz}^{\mathrm{Ge(2)}}$\,$\parallel$\,$ab$; see the SM~\cite{SM}).

\begin{figure}[t]
	\includegraphics[width=7.6cm]{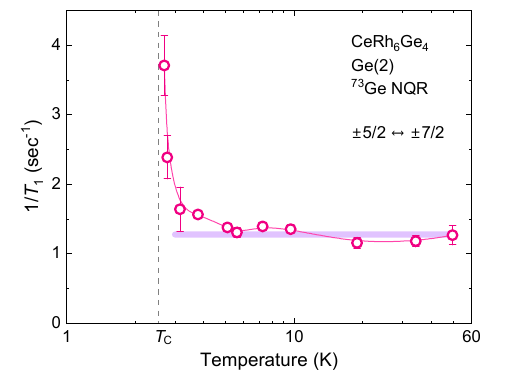}
	\vskip -0.25cm
	\caption{Zero-field $^{73}$Ge NQR $1/T_1$ of CeRh$_6$Ge$_4$ at Ge(2) plotted as a function of temperature, determined from the $I_z = \pm 5/2 \leftrightarrow \pm 7/2$ line (see the SM~\cite{SM} for details). The bold horizontal line is a guide to the eyes, and the vertical dashed line indicates the FM transition temperature $T_\mathrm{C}$ ($=$\,2.51\,K).}
	\label{fig:T1}
\end{figure}

The magnetic fluctuations in the PM state ($T>T_\mathrm{C}$) were investigated with zero-field $^{73}$Ge NQR by measuring the nuclear spin-lattice relaxation rate $1/T_1$ that probes the dynamic spin susceptibility at Ge sites. In the most generic case, it is expressed as~\cite{Moriya1963}
\begin{equation}
	\frac{1}{T_1}= \frac{\gamma^2_nk_\mathrm{B}T}{\gamma^2_e\hbar^2}\sum_q A_\perp^2(q)\frac{\chi^{''}_{\perp}(q,\omega_N)}{\omega_N},
 \label{eq:T1}
 \tag{2}
\end{equation}
where $\chi''_{\perp}(q,\omega)$ is the imaginary part of the local dynamic spin susceptibility normal to the the $V_{zz}$ direction, $q$ is the wavenumber, $A_\perp$ is the transverse component of the hyperfine coupling tensor, $\gamma_e$ is the electronic gyromagnetic ratio, and $\omega_N$ is the nuclear Larmor frequency.
Figure~\ref{fig:T1} presents the measured $1/T_1$ plotted as a function of temperature, determined for the $I_z = \pm5/2 \leftrightarrow \pm 7/2$ transition of Ge(2).
Reasonably good fits were obtained to the relaxation curves at all measured temperatures (from which $T_1$ is derived) by assuming a standard formula for this transition and $\eta^\mathrm{Ge(2)} = 0.365$~\cite{Chepin_1991} %, which guarantees that the relaxation is due to magnetic fluctuations
(see the SM~\cite{SM} for details).   
Notably, we find a sharp increase in $1/T_1$ below $\sim$ 4.0\,K that keeps growing towards $T_\mathrm{C}$. 
The intensity and linewidth of the NQR signal in this temperature range did not show a notable change, which confirms that all nuclear spins are probed throughout the experiment and suggests that there is no (magnetic or charge) inhomogeneity developing with cooling (see the SM~\cite{SM}). 
The diverging feature in $1/T_1$ is then understood as an indication of critical slowing down representative of a %second-order magnetic 
phase transition in a clean system. 

Above 4.0\,K, temperature-independent behavior was observed in $1/T_1$ at least up to 50\,K. Behavior of this kind is often reported in 4$f$-electron materials at high temperatures, where the relaxation is predominantly caused by the PM fluctuations of 4$f$ moments~\cite{Moriya1, FRADIN, MacLaughlin_PRB, Kawasaki2008, Dioguardi_PRB2017}. 
In fact, for a $q$-independent dynamic spin susceptibility, the relaxation rate associated with such local-moment fluctuations can be expressed as~\cite{Moriya1956}
\begin{equation}
	\left(\frac{1}{T_1}\right)_{\mathrm{loc}}=\sqrt{2\pi}\left(\frac{\gamma_n A}{z'}\right)^2\frac{z'\mu_\mathrm{eff}^2}{3\omega_\mathrm{ex}}
 \label{eq:LMF}
 \tag{3}
\end{equation} 
with the exchange frequency $\omega_\mathrm{ex}$ given by
\begin{equation}
	(\hbar\omega_\mathrm{ex})^2=(k_\mathrm{B}\Theta)^2\frac{3g_J}{z|g_J-1|^3\mu_\mathrm{eff}^2},
 \label{eq:exchange}
 \tag{4}
\end{equation}

\noindent where $z$ is the number of nearest-neighbor magnetic ions around the target nucleus, $z'$ is the number of magnetic ions coupling to that nucleus [$z=z'=2$ for the case of Ge(2)], $\Theta$ is the Curie-Weiss temperature, $g_J$ is the Land\'{e} $g$ factor, and $\mu_\mathrm{eff}$ is the effective moment. 
For CeRh$_6$Ge$_4$, two low-lying CEF doublet states separated by $\approx$ 5.8\,meV~\cite{Shu} need to be considered here, which brings complication and non-negligible uncertainties in these parameters (in particular, as a function of $T$), making a quantitative estimate of $(1/T_1)_{\mathrm{loc}}$ challenging. Nonetheless, the observed nearly temperature-independent $1/T_1$ and the absence of Korringa behavior ($1/T_1 \propto T $)~\cite{Abragam1961} that is characteristic of itinerant magnets \mh{(or a heavy-fermion state in Kondo-lattice systems)} suggest a more localized character of 4$f$ electrons in the PM state.

\begin{figure*}[t]
	\includegraphics[width=\textwidth]{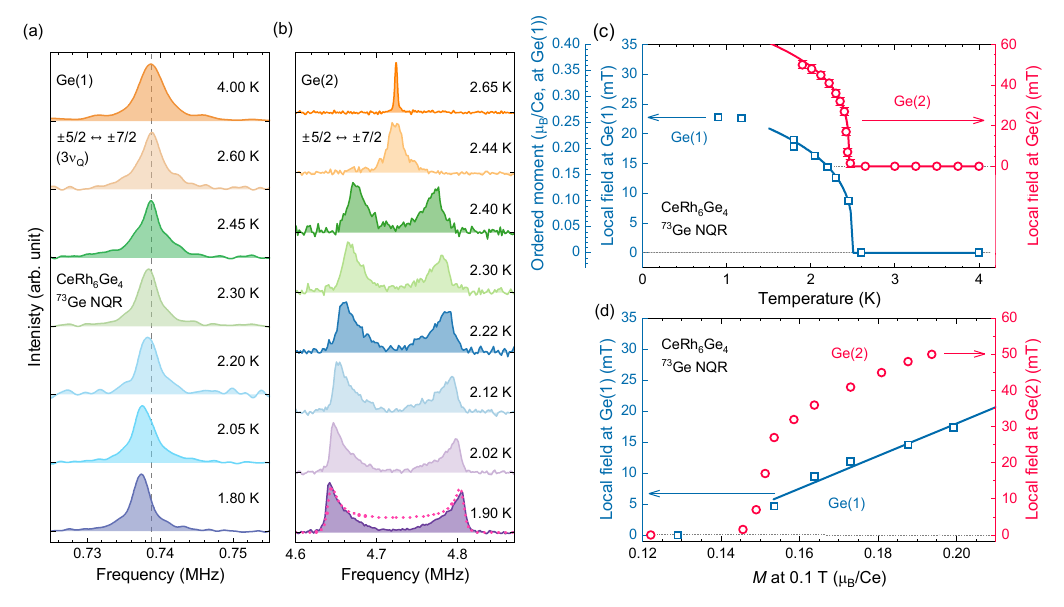}
	\caption{
    Temperature dependence of Zeeman-perturbed $^{73}$Ge NQR spectra of CeRh$_6$Ge$_4$ across $T_\mathrm{C}$ ($=$\,2.51\,K) for (a) Ge(1) and (b) Ge(2) at the $I_z=\pm5/2 \leftrightarrow \pm7/2$ line. 
The dotted curve in (b) shown on top of the data at 1.90~K represents the calculated powder pattern for the case of an isotropic angular distribution of the internal local field within the $ab$-plane and $V_{zz}^{\mathrm{Ge(2)}}$\,$\parallel$\,$ab$ (see the text for detail).  
(c) Temperature dependence of internal local fields at Ge(1), $H_\mathrm{int}^{\mathrm{Ge(1)}}$, and Ge(2), $H_\mathrm{int}^{\mathrm{Ge(2)}}$. The corresponding ordered moment size $m$ per Ce is also shown derived at Ge(1) from $H_\mathrm{int}^{\mathrm{Ge(1)}}$ and the magnetization $M$ data in (d) using the hyperfine coupling constant $A_\mathrm{FM}^{\mathrm{Ge(1)}}$ determined by the fit in (d) and extrapolated to low temperature as $m=H^{\mathrm{Ge(1)}}_\mathrm{int}/A_\mathrm{FM}^{\mathrm{Ge(1)}}$.
Solid curves are fits to the data with $M_0(1-T/T_\mathrm{C})^\beta$ by fixing the value of $T_\mathrm{C} $ to 2.51~K. 
(d) $H_\mathrm{int}^{\mathrm{Ge(1)}}$ and $H_\mathrm{int}^{\mathrm{Ge(2)}}$ plotted against the magnetization $M$ (at 0.1~T $\parallel$\,$ab$) for the temperature range between 1.9 and 2.6~K. 
The line gives a linear fit to the Ge(1) data, the slope of which determines $A_\mathrm{FM}^{\mathrm{Ge(1)}}=87.5 \pm 9.5$\,mT/$\mu_\mathrm{B}$. } 

	\label{fig:Moment}
\end{figure*}

Next, to gain insight into the microscopic nature of magnetic order, zero-field $^{73}$Ge NQR measurements were extended down to 0.90\,K. 
A clear change was found in the pulse conditions at $T<T_\mathrm{C}$ relative to what were used in the PM state. 
This is most likely due to domain formation in the ordered state and the associated amplification of rf pulses by the ordered moments both inside the domains and at their boundaries, as often observed in ferromagnetic and ferrimagnetic materials in their ordered states~\cite{turov1972nuclear}. 
Relatively small changes were observed in the spectra below $T_\mathrm{C}$ for both Ge(1) and Ge(2), suggesting that the internal field due to order is rather small, which agrees with an earlier $\mu$SR report~\cite{Shu}.
The dominant contribution to the spectrum thus remains the quadrupole term in Eq.~(\ref{eq_NQR_Zeeman}), and the Zeeman term gives only a perturbative correction (this situation is known as Zeeman-perturbed NQR~\cite{Kitaoka1987}).   

Figures~\ref{fig:Moment}(a) and \ref{fig:Moment}(b) show representative NQR spectra for Ge(1) and Ge(2) measured across $T_\mathrm{C}$~for the $I_z=\pm5/2 \leftrightarrow \pm7/2$ line.  
For axially-symmetric Ge(1) ($\eta =0$), the ($3 \nu_{\mathrm{Q}}$) line shape stays qualitatively unchanged with cooling, but a very small frequency shift, $\Delta \nu$, without an excess broadening is observed below $T_\mathrm{C}$ of a size of $(\Delta \nu/\nu_0)^{\mathrm{Ge(1)}} \approx -0.3\,\%$ at 1.8~K (with $\nu_0 = 3 \nu_{\mathrm{Q}}=$~0.739\,MHz) [Fig.~\ref{fig:Moment}(a)]. A similarly minor shift without splitting is found for the $I_z=\pm7/2 \leftrightarrow \pm9/2$ (4$\nu_{\mathrm{Q}}$) line, while a small but finite splitting appears for the $I_z=\pm3/2 \leftrightarrow \pm5/2$ (2$\nu_{\mathrm{Q}}$) line (see the SM~\cite{SM}). 
In general, the internal local field at Ge(1) due to magnetic order, $\textbf{H}_{\mathrm{int}}^{\mathrm{Ge(1)}}$, can lift the spin degeneracy of $I_z$ states through nuclear Zeeman interaction and within the first-order perturbation framework for 2$\nu_{\mathrm{Q}}$, 3$\nu_{\mathrm{Q}}$, and 4$\nu_{\mathrm{Q}}$ lines, cause a line splitting that scales to $\propto \gamma_n H_{\mathrm{int}}^{\mathrm{Ge(1)}}\mathrm{cos}~\theta$~\cite{Abragam1961}, where $\theta$ is the angle between $\textbf{H}_{\mathrm{int}}^{\mathrm{Ge(1)}}$ and $V_{zz}^{\mathrm{Ge(1)}}$ ($\parallel$\,$c$) and $H_{\mathrm{int}}^{\mathrm{Ge(1)}} = \vert\textbf{H}_{\mathrm{int}}^{\mathrm{Ge(1)}}\vert$. 
The absence of any splittings at $3\nu_{\mathrm{Q}}$ and $4\nu_{\mathrm{Q}}$ lines suggests $\theta = 90^{\circ}$, which means that the local field at Ge(1) points perpendicular to the $c$-axis.  
The small frequency shifts at these lines and the small splitting at 2$\nu_{\mathrm{Q}}$ line are, however, not expected within the first-order framework, suggesting the presence of higher-order effects. 
To take into account such effects, we have performed exact diagonalization of Eq.~(\ref{eq_NQR_Zeeman}) and simultaneously fit three Ge(1) lines at 2.0\,K.  
This model reproduces the data excellently (see the SM~\cite{SM} for details), which reveals that both $\nu_{\mathrm{Q}}^\mathrm{Ge(1)}$ and $\eta^\mathrm{Ge(1)}$ do not vary across $T_\mathrm{C}$ and all of the spectral changes can be accounted for by a single internal field that lies exactly within the $ab$-plane ($\textbf{H}_{\mathrm{int}}^{\mathrm{Ge(1)}}$\,$\parallel$\,$ab$). 
Furthermore, by considering the local site symmetry and the hyperfine coupling tensor $\mathbb{A}$ that connects the ordered moment $\textbf{m}$ to the local field, $\textbf{H}_{\mathrm{int}}^{\mathrm{Ge(1)}} =\mathbb{A} \cdot\textbf{m}$, we conclude that the moment $\textbf{m}$ is confined to the $ab$-plane ($\textbf{m}$\,$\parallel$\,$ab$) without any spatial modulation in its size (for details, see the SM~\cite{SM}). 

In contrast to Ge(1), the spectrum of Ge(2) for the $I_z=\pm5/2\,\leftrightarrow\,\pm7/2$ transition shows a notable line splitting and a spectral broadening below $T_\mathrm{C}$, with a peak-to-peak separation of the spectrum,  $\delta \nu$, growing upon cooling and its center of gravity position remaining unaltered [Fig.~\ref{fig:Moment}(b)]. 
The spectrum at 2.44\,K exhibits a transient feature without a clear splitting but a pedestal-like tail, which might be a sign of phase separation due to weakly first-order nature of the FM transition. A phase-separation-like feature is also reported in Zeeman-perturbed $^{59}$Co NQR in UCoGe~\cite{hattori201159co,hattori2010weakly}.
The size of splitting reaches $(\delta \nu/\nu_0)^{\mathrm{Ge(2)}}$\,$\approx$\,$4.3\,\%$ at 1.9\,K (with $\nu_0=$~4.723\,MHz). A splitting of a similar size is also observed for the $I_z=\pm7/2 \leftrightarrow \pm9/2$ line (see the SM~\cite{SM}). 
Interestingly, the Ge(2) spectrum shows a powder-pattern-like feature below $T_\mathrm{C}$ with the inner tails of the double horn spreading more towards the center where it has vanishingly small intensities, whereas the outer tails are steeper. % more steep and quickly cut off. 
The appearance of such a pattern indicates that there is a distribution in the direction of the internal local field at Ge(2), $\textbf{H}_{\mathrm{int}}^{\mathrm{Ge(2)}}$, relative to that of $V_{zz}^{\mathrm{Ge(2)}}$;
the histogram of this distribution should be reflected in the pattern. 
From the condition of $\textbf{m}$\,$\parallel$\,$ab$ derived from the Ge(1) spectral data and by considering the local-site symmetry at Ge(2), we show that $\textbf{H}_{\mathrm{int}}^{\mathrm{Ge(2)}}$\,$\parallel$\,$ab$ holds (see the SM for details~\cite{SM}). 

Then, following the conclusion of the NMR experiments that gives $V_{zz}^{\mathrm{Ge(2)}}$\,$\parallel$\,$ab$ (see Fig.~\ref{fig:Ge2-NMR-out-of-plane-field}%see the SM~\cite{SM}
), we calculated the spectral pattern for Ge(2) with exact diagonalization of Eq.~(\ref{eq_NQR_Zeeman}) and sought %for 
the %best 
distribution of $\textbf{H}_{\mathrm{int}}^{\mathrm{Ge(2)}}$ within the $ab$-plane in agreement with experiment. 
The dashed curve at the bottom of Fig.~\ref{fig:Moment}(b) presents the best agreement %result 
for that obtained by assuming an isotropic angular distribution of $\textbf{H}_{\mathrm{int}}^{\mathrm{Ge(2)}}$ within the $ab$-plane. 
The calculation accounts for the primary features of the spectrum including the sharp edges and the stretched feature of the horns extending to the center. A similarly good agreement is achieved with the same parameters for the $I_z=\pm7/2 \leftrightarrow \pm9/2$ line (see the SM~\cite{SM}). 
Given that we use powdered samples, the above agreement naturally suggests that the ordered moment $\textbf{m}$ does not have a preferential direction of order inside the $ab$-plane. 
The direction of $\textbf{m}$ varies between different domains and crystallites and is distributed randomly within the plane, resulting in the powder-like spectral pattern. 
Therefore, the analyses of the Ge(1) and Ge(2) spectra together provide evidence that the easy-plane FM order in CeRh$_6$Ge$_4$ has an isotropic (XY-type) character. 

Around the center of the Ge(2) spectrum, however, the agreement between the calculation and experiment is poor; namely, the former has %a descent 
intensity around the center, whereas the latter has almost none. 
The calculated distribution of the field orientation reveals that the two horns in the spectrum correspond to nuclear spins that satisfy $\textbf{H}_{\mathrm{int}}^{\mathrm{Ge(2)}}$\,$\parallel$\,$V_{zz}^{\mathrm{Ge(2)}}$, while the intensity around the center arises from those spins approximately fulfilling $\textbf{H}_{\mathrm{int}}^{\mathrm{Ge(2)}}$\,$\perp$\,$V_{zz}^{\mathrm{Ge(2)}}$. 
The observed discrepancy at the center of the spectrum thus indicates that the spectral weight for the spins under transverse fields (perpendicular to $V_{zz}^{\mathrm{Ge(2)}}$) is greatly suppressed relative to those feeling longitudinal fields (parallel to $V_{zz}^{\mathrm{Ge(2)}}$). \mh{(Note that this suppression is not related to a distribution of the spin-spin relaxation time $T_2$ across the spectrum; $T_2$ was indeed frequency independent.)}

For zero-field NMR in ferromagnets, the external rf field, $H_1$, generally induces fluctuations of the ordered moment $\textbf{m}$ that causes a transverse internal rf field, $h_\perp$, at the nuclear positions perpendicular to their quantization axis, which is known to drastically enhance the NMR signal intensity through the hyperfine coupling, characterized by an enhancement factor $\epsilon = (h_\perp / H_1) ^2$ \cite{turov1972nuclear}.
When the Zeeman term in Eq.~(\ref{eq_NQR_Zeeman}) is much larger than the quadrupole term ($\mathcal{H}_{Z} \gg \mathcal{H}_{Q}$) as in standard zero-field NMR\,\cite{Gossard_PRL,Portis}, the quantization axis is along the direction of $\textbf{H}_{\mathrm{int}}$. 
By contrast, if the quadrupole term is dominant as in our Zeeman-perturbed NQR ($\mathcal{H}_{Z} \ll \mathcal{H}_{Q}$), it is reasonably assumed to be parallel to $V_{zz}$. 
For either case, signal enhancement is expected to occur, with the value of $\epsilon$ depending on the distribution and size of  $h_\perp$\,\cite{Stearns_1967}. 
We speculate that the remarkable depression of the Ge(2) NQR spectrum near its center would be due to a spread of $\epsilon$ across the spectrum in a way such that the enhancement is preferentially suppressed towards the center and, for the spins feeling $\textbf{H}_{\mathrm{int}}^{\mathrm{Ge(2)}}$\,$\perp$\,$V_{zz}^{\mathrm{Ge(2)}}$,  becomes vanishingly small.  
The internal rf field felt by such spins cannot have large transverse components normal to $V_{zz} ^{\mathrm{Ge(2)}}$ within the $ab$-plane (i.e., $h_\perp^{ab} \approx 0$), as in-plane fluctuations of $\textbf{m}$ only cause longitudinal rf fields for these spins.  
Consequently, to amplify the signal, a transverse rf field that is perpendicular to the plane, $h_\perp^c$, is necessary, which by symmetry can be induced only through out-of-plane fluctuations of $\textbf{m}$. 
The depression of the spectrum would therefore indicate that there is a magnetic stiffness in the ordered moment that forces the magnetization to stay within the $ab$-plane and results in $h_\perp^c \approx 0$.

Figure~\ref{fig:Moment}(c) presents the largest internal local field at Ge(1) and Ge(2) versus temperature determined from the shift $\Delta \nu$ and splitting $\delta \nu$
in Figs.~\ref{fig:Moment}(a) and \ref{fig:Moment}(b).  Conventional order parameter-like temperature dependence is found below $T_\mathrm{C}$ with a saturated feature below $\sim$\,1\,K. 
Fitting them with a form $M_0(1-T/T_\mathrm{C})^\beta$, critical exponents are obtained as $\beta = 0.304\pm0.010$ for Ge(1) and $\beta = 0.280\pm0.024$ for Ge(2). These values of $\beta$ are quite different from the mean-field value ($\beta = 0.5$)~\cite{Guillou1980}. 
In Fig.~\ref{fig:Moment}(d) the local fields are plotted relative to the bulk magnetization $M$ measured in the same sample between 1.9 and 2.6\,K.
A clear linear relationship between the two quantities is seen below $T_\mathrm{C}$, which indicates that the transferred hyperfine interaction---of the Ce 4$f$ moments with the ${}^{73}$Ge nuclei---is responsible for the internal local fields.
Furthermore, our simulation of the direct dipole field at the Ge sites due to Ce moments is more than a factor of five too small to account for the observed hyperfine fields (see the SM~\cite{SM} for details). 
By performing a linear fit to the $M$--$H^{\mathrm{Ge(1)}}$ plot, the in-plane component of the hyperfine coupling constant is determined as $A_\mathrm{FM}^{\mathrm{Ge(1)}}=87.5 \pm 9.5$\,mT/$\mu_\mathrm{B}$ at Ge(1) per nearest-neighbor Ce atoms.  [Estimating the coupling at Ge(2) is difficult due to lower local-site symmetry.]

Using this coupling constant, the size of the ordered moment, $m=H^{\mathrm{Ge(1)}}_\mathrm{int}/A_\mathrm{FM}^{\mathrm{Ge(1)}}$, can be extrapolated down to the lowest measured temperature, which yields a saturated value of $m\,\approx\,0.26\pm0.03$\,$\mu_\mathrm{B}$/Ce at 0.90\,K [Fig.~\ref{fig:Moment}(c)], in good agreement with earlier magnetization measurements at 0.44\,K reporting 0.2\,--\,0.3\,$\mu_\mathrm{B}$~\cite{Shen} and with an estimate from $\mu$SR \cite{Shu}. 
As mentioned earlier, the ordered moment size in CeRh$_6$Ge$_4$ is much smaller than what is expected for the corresponding size of the $\Gamma_7$ ground-state CEF doublet of 1.28\,$\mu_\mathrm{B}$~\cite{Shen}. Given that an XY-type easy-plane anisotropy is found in our NQR study, this suppressed moment size could be an indication of strong magnetic frustration anticipated in easy-plane FM systems with a Q1D %quasi-one-dimensional
nature~\cite{Komijani_PRL} or due to Kondo screening.

\begin{figure}[t]
	\includegraphics[width=8.6cm]{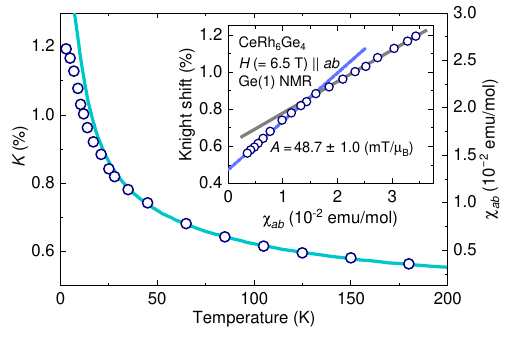}
	\vskip -0.25cm
	\caption{$^{73}$Ge NMR in-plane Knight shift $K$ of CeRh$_6$Ge$_4$ at Ge(1) (symbols) and in-plane DC magnetic susceptibility $\chi_{ab} = M_{ab}/H$ (solid curve) plotted as a function of temperature at $H$\,$=$\,6.5\,T ($||~ab$). Inset: $K$ plotted against $\chi_{ab}$ with temperature used as an implicit parameter. Lines stand for the least square fits to high and low-temperature parts. A slope change (intersection of the two fit lines) appears at $T^*$\,$\approx$\,25\,K.} 
	\label{fig:NMR}
\end{figure}

Lastly, we look at the NMR Knight shift $K$ in the PM state. 
Figure~\ref{fig:NMR} shows the temperature dependence of in-plane $K$ ($H$\,$\parallel$\,$ab$) at Ge(1) recorded at $H=$ 6.5\,T for the central line (see Fig.~\ref{fig:NMR_full_spectrum}). 
The corresponding in-plane DC magnetic susceptibility, $\chi_{ab} = M_{ab}/H$, 
measured at the same field is plotted together in the figure, which shows Curie-Weiss behavior in agreement with earlier studies at lower field \cite{Matsuoka}. %(see the SM\cite{SM} for details). 
A linear scaling is seen between $K$ and $\chi_{ab}$ above $T^* \approx$ 25\,K (the inset of Fig.~\ref{fig:NMR}). 
Making a linear fit to this $K$--$\chi_{ab}$ plot with a form $K=(A/N_{\mathrm{A}}\mu_\mathrm{B}) \chi_{ab}$ yields $A = 48.7$\,mT$/\mu_{\mathrm{B}}$ ($T > T^*$), where $A$ is the transferred hyperfine coupling constant between Ge(1) and Ce 4$f$ moment and $N_{\mathrm{A}}$ is the Avogadro's number. 
A smaller hyperfine coupling constant in the PM state is also
reported by $^{69}$Ga NMR in the uranium-based itinerant ferromagnet UGa$_2$ \cite{Kambe_2014}.
There is a slight slope change at $T^*$, as reflected in a kink in the $K$--$\chi_{ab}$ plot, and another linear regime is realized below $T^*$, with a slightly smaller hyperfine coupling value of $A= 32.2$\,mT/$\mu_{\mathrm{B}}$ ($T < T^*$).  
A kink in the $K$--$\chi$ relation is widely reported in Kondo-lattice systems, in which the anomaly is often associated with (i) local magnetic inhomogeneity~\cite{MacLaughlin_PRB, Bordelon2025}, developing Kondo coherence~\cite{Kim_PRL,Yang_PRL,Yang_2008}, or CEF level-mixing effects~\cite{Ohama_JPSJ,Curro_PRB,Ohishi_PRB}.
%The 
Inhomogeneity (i) is unlikely here, as our NMR linewidth follows the bulk susceptibility as expected for clean uniform systems (see the SM~\cite{SM}).  
The shift anomaly is then suggestive of either Kondo coherence (ii) or CEF mixing (iii) or both. 

In CeRh$_6$Ge$_4$ the energy scales characteristic of these two phenomena are not clear. % could be rather close to $T^*$. 
Earlier studies argued that the Kondo temperature scale is $T_\mathrm{K} \sim 19$\,K from entropy analysis of specific heat assuming a spin-1/2 Kondo model~\cite{Matsuoka}, while the resistivity measurements suggested $T_\mathrm{K} \sim 130$\,K from the peak in the temperature dependence of its magnetic contribution~\cite{Matsuoka}. 
A later susceptibility measurement suggested that a first-excited-state CEF doublet locates 5.8~meV ($\sim$ 67~K) above the ground-state CEF doublet~\cite{Shu}.
However, inelastic neutron scattering experiments found a broad excitation with a full width at half maximum of $\sim$\,30\,meV, which implies a Kondo scale with an enhanced degeneracy (meaning that the Kondo scale is much larger than the CEF level separation)~\cite{Shu}. 
This last point is further supported by the Kadowaki-Woods ratio that points to a four-fold-degenerate ground state~\cite{Shen}.
The shift anomaly seen by NMR could, therefore, be due to a complex interplay of Kondo and CEF effects in this compound. 

Naively, the presence of dominantly local fluctuations as inferred from our temperature dependent $1/T_1$ measurements from 5\,--\,50\,K contradicts our shift anomaly seen at $T^*$\,$\approx$\,25\,K and the expectation of strong Kondo screening that would be inferred from the large energy scale observed by inelastic neutron scattering measurements~\cite{Shu}. 
A resolution to this might be found in the anisotropic hybridization to different CEF levels as has been suggested in previous studies~\cite{Wu_ARPEX, Pei}.

\section{Summary}
We have performed $^{73}$Ge nuclear quadrupole resonance (NQR) and magnetic resonance (NMR) spectroscopy in the quantum critical ferromagnet CeRh$_6$Ge$_4$ at ambient pressure. 
Our NQR and NMR measurements in the paramagnetic state reveal two Ge resonances and determine their electric field gradient tensors as well as the directions of their principal axes relative to the hexagonal basal plane. 
The nuclear spin-lattice relaxation rate $1/T_1$ measurements find a critical slowing down near $T_\mathrm{C}$ characteristic of a ferromagnetic phase transition and spin fluctuations predominantly linked to local-moment fluctuations. 
The Knight shift measurements find Curie-Weiss behavior at high temperature and a deviation from that below $T^*$\,$\approx$\,25\,K possibly due to a mixture of CEF level depopulation effects and Kondo screening.  
Zeeman-perturbed NQR measurements in the ordered state reveal a uniform ferromagnetic order with a small ordered moment of $\approx$\ 0.26\,$\mu_\mathrm{B}$/Ce at Ge(1) that is pinned within the $ab$-plane. 
The ordered moment shows a magnetic in-plane stiffness against out-of-plane radiofrequency fields and has an isotropic (XY-type) easy-plane character.
Our microscopic studies of %local 
magnetism in this compound provide new insight into %local 
magnetic anisotropy and fluctuations in CeRh$_6$Ge$_4$ that would be important for designing a more realistic theoretical framework to describe the quantum criticality in this compound under pressure. \\

\begin{acknowledgments}
	We thank H. Sakai, S. Lin, P. F. S. Rosa, S. M. Thomas, S. E. Brown, and N. J. Curro for fruitful discussions. NMR, NQR, and thermodynamic measurements were supported by the U.S. Department of Energy, Office of Basic Energy Sciences, Division of Materials Science and Engineering project ``Quantum Fluctuations in Narrow-band Systems''. Crystal synthesis and basic characterization were supported by Laboratory Directed Research and Development Program, Exploratory Research project 20240109ER.
\end{acknowledgments}

\end{document}

\renewcommand{\thetable}{S\arabic{table}}
\renewcommand{\thefigure}{S\arabic{figure}}

\title{Supplemental Material for ``Evidence for easy-plane XY ferromagnetism in heavy-fermion quantum-critical CeRh$_6$Ge$_4$"}

\author{Riku Yamamoto$^{1}$}
\author{Sejun Park$^1$}
\author{Zachary W. Riedel$^{1}$}
\author{Phurba Sherpa$^{1,2}$}
\author{Joe D. Thompson$^{1}$}
\author{Filip Ronning$^{1}$}
\author{Eric D. Bauer$^{1}$}
\author{Adam P. Dioguardi$^{1}$}
\author{Michihiro Hirata$^{1}$}

\affiliation{$^{1}$ Materials Physics and Applications $-$ Quantum, Los Alamos National Laboratory, Los Alamos, New Mexico  87545, USA}
\affiliation{$^2$ Department of Physics and Astronomy, University of California, Davis, California 95616, USA}

\date{\today}

\maketitle

\section{Sample growth and characterization}

\subsection{Synthesis and crystal structure}
\begin{figure}[b]
	\includegraphics[width=8.6cm]{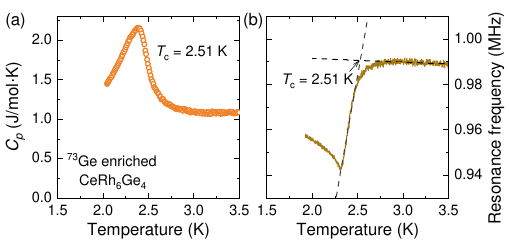}
	%\vskip -0.25cm
	\caption{Specific heat $C_p$ (a) and resonance frequency of the LC tank circuit (b) in $^{73}$Ge-labeled CeRh$_6$Ge$_4$ samples, plotted as a function of temperature. Dashed lines in (b) give two fitting lines to the data below and above the ferromagnetic transition temperature of $T_\mathrm{C} =$~2.51~K.}
	\label{fig:intro}
\end{figure}

Single-crystalline, polycrystalline, and powder samples of CeRh$_6$Ge$_4$ were synthesized from Bi flux~\cite{Voßwinke}.
The single crystal and polycrystalline samples were used in DC magnetic susceptibility and specific heat measurements, respectively, while the powder (or a mosaic of single-crystalline) samples were used in NMR, NQR, and AC magnetic susceptibility %, and specific heat
measurements. 
The Ge atoms in these samples were selectively replaced with an enriched $^{73}$Ge isotope (nuclear spin $I$\,$=$\,$9/2$) by 95.60\,\% for $^{73}$Ge NQR and NMR to gain signal intensities. 
The $^{73}$Ge-enrichment did not affect the magnetic transition relative to that in as-grown samples~\cite{Kotegawa, Shen} (see Fig.~\ref{fig:intro}). 

CeRh$_6$Ge$_4$ crystallizes in a hexagonal structure (with space group $P{\bar6m2}$, No.~187), as illustrated in Fig.~1(a) in the main text.
There are two crystallographically nonequivalent Ge sites in the unit cell, dubbed $j=$~Ge(1) and Ge(2). 
The Ge(1) atoms sit at the center of the triangle lattice formed by surrounding Ce atoms within the same basal plane and have an axial local-site symmetry (Wyckoff letter $1c$ and point group symmetry $\bar6m2$). 
The other Ge(2) atoms form an array of triangles in another plane shifted by $c/2$ out of the Ce triangle-lattice plane and have lower local site symmetry ($3k$ and $mm2$).

\subsection{Heat capacity and DC and AC magnetic susceptibility measurements}

Magnetization and DC magnetic susceptibility were measured using a Quantum Design Magnetic Property Measurement System (MPMS) in the temperature ($T$) range between 1.8 to 300\,K and magnetic field ($H$) up to 6.5\,T for an in-plane field orientation ($H~||~ab$). 
Zero-field heat capacity was measured with a Quantum Design Physical Properties Measurement System (PPMS). % from 2.0 to 250~K.
AC magnetic susceptibility was measured at zero field using a solenoid coil in an inductor-capacitor (LC) circuit between 1.9 to 4.0\,K. 
The resonance frequency of this tank circuit is given by $f= 1/2\pi\sqrt{LC}$, with the inductance $L$ and capacitance $C$. 
The coil acts as an inductor that is filled with powder samples up to a filling factor of $P \sim$~90~\%. The shift of the resonance frequency, $\Delta f$, is connected to the AC magnetic susceptibility of the sample, $\chi_\mathrm{ac}$, through
\begin{equation}
	\frac{\Delta f}{f_0} = -\frac{P}{2}~4\pi(\chi_\mathrm{ac}+\chi_\mathrm{skin}),
 \label{eq_ac_chi}
\end{equation}  
\noindent where $f_0$ is a reference frequency, $\chi_\mathrm{ac}$ is connected to the magnetization $M$ via $M=\chi_\mathrm{ac}H$, and $\chi_\mathrm{skin}$ is a diamagnetic signal coming from the skin effect in normal metals and is linked to the skin depth $\delta=\sqrt{2 \rho/\mu f}$, with the permeability of the sample $\mu$~\cite{Vannette2008}. 
Using the reported resistivity value in this temperature range~\cite{Shen}, the frequency we use $f_0 \approx$~0.99\,MHz gives a skin depth of $\delta \approx$~30\,$\mu$m, which is an order of magnitude larger than the grain size of our samples, $d \ll\delta$. 
The diamagnetic term in Eq.~(\ref{eq_ac_chi}) is negligibly small in this limit  \cite{Hardy1993}, and the dominant contribution to $\Delta f$ comes from $\chi_\mathrm{ac}$ in the vicinity of the ferromagnetic transition.

The transition temperature in our samples was $T_\mathrm{C} =$\,2.51\,K, as determined from the midpoint of the jump in the heat capacity $C_p$ [Fig.~\ref{fig:intro}(a)] and the intersection of the linear fits to $\Delta f$ at low-$T$ (2.3\,K $\leq T \leq$ 2.6\,K) and high-$T$ (3.0\,K~$\leq T \leq$ 3.5\,K) regions [dashed lines in Fig.~\ref{fig:intro}(b)].

\section{Germanium-73 NMR and NQR measurements}
\subsection{Experimental details}

Powdered CeRh$_6$Ge$_4$ samples with a crystallite size smaller than 2\,$\mu$m were loaded in plastic tubing, capped with quartz wool, and inserted into a copper solenoid coil for NMR and NQR experiments. Two titanium frits with 2\,$\mu$m diameter pores were installed at both ends of the tubing to seal the containment and to allow thermal contact with $^4$He exchange gas or liquid in the cryostat. 

NMR and NQR measurements were performed at $^{73}$Ge nuclei (gyromagnetic ratio $^{73}\gamma_n/2\pi$ = 1.4897\,MHz/T, quadrupole moment $^{73}Q = -0.196$~barn~\cite{Harris}) using standard spin-echo and Carr-Purcell-Meiboom-Gill (CPMG) pulse sequence techniques \cite{CP,MG}  and commercially available spectrometers (REDSTONE, Tecmag Inc.).  
Typical pulse widths used were $t_{\pi/2} = 10$--$40\,\mu$s, with an inter-pulse spacing (between excitation $\pi/2$ pulse and refocusing $\pi$ pulse) being $\tau = 500$--$1000\,\mu$s. 
The spectra were obtained from an echo signal recorded at a fixed carrier frequency through the standard Fast Fourier Transform (FFT).

\begin{figure}[b]
	\includegraphics[width=8.6cm]{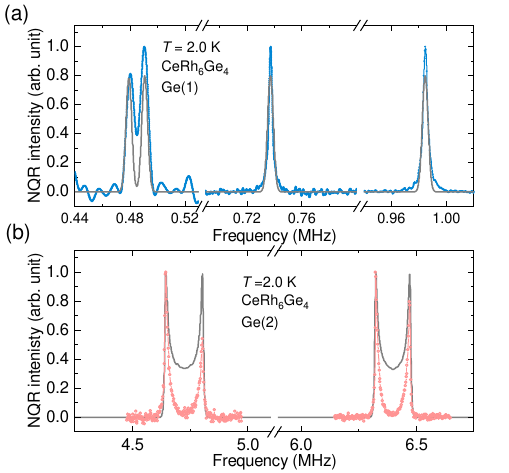}
	%\vskip -0.25cm
	\caption{Ordered-state zero-field $^{73}$Ge NQR Spectrum at (a) Ge(1) (blue) and (b) Ge(2) (red) recorded at 2.0 K. Gray solid curves are fits to the spectrum with exact diagonalization of Eq.~(1) in the main text, with an in-plane internal local field of 19.0 and 50.0 mT at Ge(1) and Ge(2) atoms, respectively. We assumed $V_{zz}^{\mathrm{Ge(1)}}$\,$\parallel$\,c and $V_{zz}^{\mathrm{Ge(2)}}$\,$\parallel$\,$ab$. (See the main text for details.)} 
	\label{fig:SPC_FM}
\end{figure}

For $^{73}$Ge NMR measurements for $H$\,$\parallel$\,$ab$, the powdered samples were loosely packed into the tubing with a filling factor of approximately 50\,\%. 
The tubing was then cooled to 4.0\,K and subjected to a magnetic field of 12\,T to let the crystallites self-align~\cite{Farrell,Takigawa}, reflecting  the easy-plane magnetic anisotropy ($\chi_{||}/\chi_{\perp} \sim 100$ at 4.0\,K~\cite{Shen}). 
Measurements at the Ge(1) site were performed in this $H$\,$\parallel$\,$ab$ condition right after the alignment at low temperature. 
No qualitative change was observed in the spectrum up to 150\,K, which guarantees that the self-alignment holds well against temperature variation.
For $H$\,$\parallel$\,$c$, we  separately made a mosaic of small single crystals and aligned them along the $c$-axis direction. By applying the field along that direction, we measured the spectrum at the Ge(2) site. 
The full NMR spectra were recorded either by varying the field at a fixed frequency and integrating the echo files at each field in the time domain, or by varying the frequency at a fixed external field and combining the FFTs in the frequency domain. 

\begin{figure}[!t]
	\includegraphics[width=8.6cm]{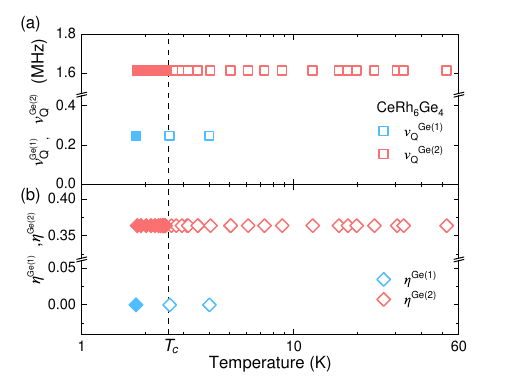}
	\vskip -0.25cm
	\caption{$^{73}$Ge NQR parameters (a) $\nu_Q$ and (b) $\eta$ plotted against temperature for Ge(1) (blue) and Ge(2) (red). Open symbols represent the data in the paramagnetic state, and filled symbols correspond to those in the ferromagnetic state. } 
	\label{fig:NuQ_eta}
\end{figure}

\begin{figure}[!t]
	\includegraphics[width=8.6cm]{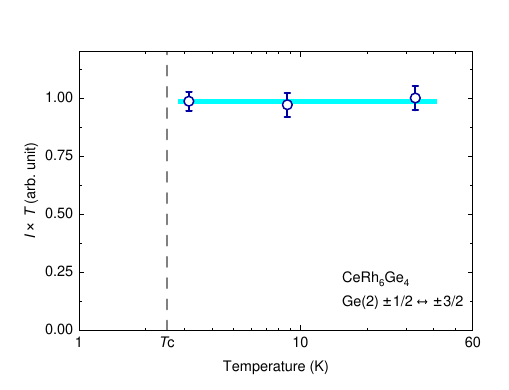}
	\vskip -0.25cm
	\caption{Temperature dependence of $^{73}$Ge NQR signal intensity times temperature above $T_\mathrm{C}$ for the $\pm$1/2 $\leftrightarrow$ $\pm$3/2 transition at Ge(2). Corrections for the $T_1$ and $T_2$ effects are reflected. Line is a guide to the eyes.} 

    \vspace{1.0cm}

	\label{fig:Intensity}
\end{figure}

$^{73}$Ge NQR spectra for Ge(1) and Ge(2) were analyzed and fit by performing exact diagonalization of Eq.~(1) in the main text using Python both in the paramagnetic state ($T>T_\mathrm{C}$) and ordered state ($T<T_\mathrm{C}$) to extract NQR parameters (i.e., quadrupole frequency $\nu_\mathrm{Q}$ and asymmetry parameter $\eta$) as well as the size of internal local magnetic field at Ge(1) and Ge(2) (Fig.~\ref{fig:SPC_FM}). 
Little temperature dependence was observed in $\nu_\mathrm{Q}$ and $\eta$ from 60~K across $T_\mathrm{C}$ (Fig.~\ref{fig:NuQ_eta}). 
The signal intensity corrected by the Boltzmann factor, $I \times T $, as well as the $T_1$ and $T_2$ effects, stayed constant down to the vicinity of $T_\mathrm{C}$ (Fig.~\ref{fig:Intensity}), which reflects that the same number of nuclear spins are probed in the experiment without any signal loss linked to electronic inhomogeneity that sometime develops in correlated electron systems with cooling~[30].

NQR spin-lattice relaxation time, $T_1$, was determined at zero field in the paramagnetic state ($T > T_\mathrm{C}$) using the inversion-recovery method, with the pulse sequence of $\pi - delay\ (t_d)- \pi/2 - \tau - \pi - echo$. 
In the vicinity of the ferromagnetic transition, a small tip angle was used for both excitation ($\pi/2$) pulses and refocusing ($\pi$) pulses~\cite{SmallTip_T1} to suppress radiofrequency heating effects.
$T_1$ was determined at the Ge(2) site for the $I_z = \pm5/2 \leftrightarrow \pm7/2$ transition from integrated echo intensity, extracted from fits to nuclear magnetization $M$ after inversion to a magnetic relaxation function valid for $I = 9/2$ and a finite asymmetry parameter $\eta$~\cite{Chepin_1991} as follows:

\newpage

\begin{equation}
	M(t_d) = M_0[1-f D^{\eta}(t_d)],
    \label{T1_relax}
\end{equation} 

\vspace{0.4cm}

\noindent where $t_d$ is the delay between the inversion and excitation pulses, $f$ is the inversion fraction, and $D^{\eta}(t_d)$ is given by

\vspace{0.2cm}

\begin{align*}
	D^{\eta}(t_d)=\ &C^\eta_{\pm \frac{1}{2}\leftrightarrow \pm\frac{3}{2}}\exp\left(-\frac{W^{\eta}_{\pm\frac{1}{2}\leftrightarrow \pm\frac{3}{2}}}{2}\frac{t_d}{T_1}\right)+\\
    &C^\eta_{\pm\frac{3}{2}\leftrightarrow \pm\frac{5}{2}}\exp\left(-\frac{W^{\eta}_{\pm\frac{3}{2}\leftrightarrow \pm\frac{5}{2}}}{2}\frac{t_d}{T_1}\right)+\\
    &C^\eta_{\pm\frac{5}{2}\leftrightarrow \pm\frac{7}{2}}\exp\left(-\frac{W^{\eta}_{\pm\frac{5}{2}\leftrightarrow \pm\frac{7}{2}}}{2}\frac{t_d}{T_1}\right)+\\
    &C^\eta_{\pm\frac{7}{2}\leftrightarrow \pm\frac{9}{2}}\exp\left(-\frac{W^{\eta}_{\pm\frac{7}{2}\leftrightarrow \pm\frac{9}{2}}}{2}\frac{t_d}{T_1}\right).
    \tag{3}
 \label{T1_9o2}
\end{align*} 

\vspace{0.4cm}

\begin{figure}[!t]
	\includegraphics[width=8.0cm]{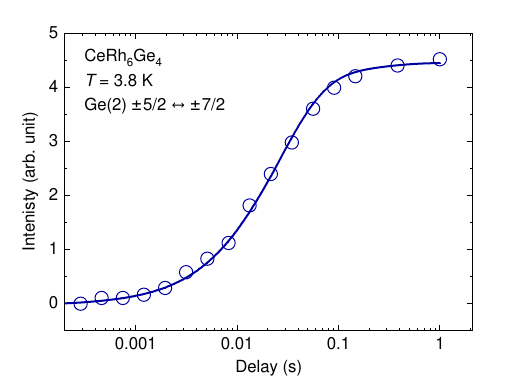}
	\vskip -0.25cm
	\caption{Representative zero-field $^{73}$Ge NQR spin-lattice ($T_1$) relaxation curve for the $I_z = \pm5/2 \leftrightarrow \pm7/2$ transition at Ge(2) recorded at 3.8~K. The solid curve is the fit with Eqs.~(\ref{T1_relax}) and (\ref{T1_9o2}) using the parameters given in Table~\ref{tab_NuQ_T1}.} 
	\label{fig:T1_Relax}
\end{figure}

\noindent Here, $C^{\eta}_{\pm(2m-1)/2 \leftrightarrow \pm(2m+1)/2}$ and $W^{\eta}_{\pm(2m-1)/2 \leftrightarrow \pm(2m+1)/2}$ ($m$ = 1, 2, 3, and 4) are the transition coefficients and exponents, respectively, that are given in Table~\ref{tab_NuQ_T1} calculated for the experimentally determined value of $\eta^{\mathrm{Ge(2)}} = 0.365$ based on our NQR spectral measurements~\cite{Chepin_1991}.
Figure \ref{fig:T1_Relax} shows a representative relaxation curve at 3.8~K fit by Eqs.~(\ref{T1_relax}) and (\ref{T1_9o2}) using the parameters in Table~\ref{tab_NuQ_T1}.

\begin{table}[b]
	\caption{Fitting parameters for $^{73}$Ge(2) spin-lattice relaxation rate $1/T_1$ for the $I_z = \pm 5/2\leftrightarrow \pm7/2$ transition with $\eta = 0.365$ \cite{Chepin_1991}}. 
    
    \vspace{0.3cm}
    
	\def\arraystretch{1.2}
	\begin{tabular}{cccc}
		\hline
		$C^\eta_{\pm \frac{1}{2} \leftrightarrow \pm \frac{3}{2}}$ & $C^\eta_{\pm \frac{3}{2} \leftrightarrow \pm \frac{5}{2}}$ & $C^\eta_{\pm \frac{5}{2} \leftrightarrow \pm \frac{7}{2}}$ & $C^\eta_{\pm \frac{7}{2} \leftrightarrow \pm \frac{9}{2}}$ \\ 
         \hline
        0.56836 & 0.37069 & 0.00134 & 0.05962 \\ \hline
        \\ \hline
		$W^\eta_{\pm \frac{1}{2} \leftrightarrow \pm \frac{3}{2}}$ & $W^\eta_{\pm \frac{3}{2} \leftrightarrow \pm \frac{5}{2}}$ & $W^\eta_{\pm \frac{5}{2} \leftrightarrow \pm \frac{7}{2}}$ & $W^\eta_{\pm \frac{7}{2} \leftrightarrow \pm \frac{9}{2}}$ \\ \hline
		60.528 & 33.994 & 17.790 & 6.110   \\ \hline	
	\end{tabular}
	\label{tab_NuQ_T1}
\end{table}

\newpage
\subsection{Transferred hyperfine coupling and local hyperfine field at Ge(1) and Ge(2)}

\begin{figure}[b]
    \vspace{0.2cm}
    \includegraphics[width=7cm]{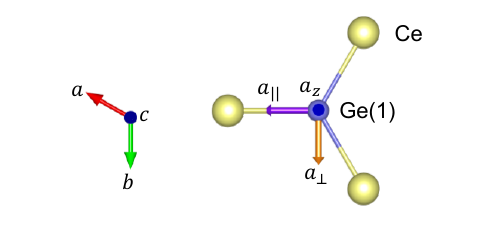}
	\vskip -0.25cm
	\caption{Illustration of the transferred-hyperfine-coupling network between nearest-neighbor Ce atoms (yellow balls) and Ge(1) atom (blue ball) defined within the hexagonal basal plane seen from the $c$-axis direction. $a_\parallel$, $a_\perp$, and $a_z$ stand for the three principal components of one of the hyperfine coupling tensors between the left-most Ce and Ge(1) defined along the $a^*$-, $b$-, and $c$-axis directions, respectively.}  
	\label{fig:CS}
\end{figure}

The Ge(1) atom in CeRh$_6$Ge$_4$ is surrounded by three nearest-neighbor (nn) Ce atoms within the hexagonal basal plane (see Fig.~1 in the main text).
Focusing on one of such three Ce-Ge(1) bonds, we can assume that the nn transferred hyperfine coupling tensor along that bond is given by a diagonal form as

\begin{equation}
\mathbb{A}^\mathrm{Ge(1)}_1 = \left[\begin{array}{ccc}
a_\parallel & 0 &  0 \\
0 & a_\perp &  0 \\
0 & 0 &  a_z
\end{array}\right],\
 \label{Ge1_tensor}
\end{equation} 

\noindent where $a_\xi$ stands for the three coupling constants along three directions [$\xi\ =(\parallel,\ \perp$, and $z)$] as defined in Fig.~\ref{fig:CS}. 
Given that there is a three-fold rotational symmetry axis at the Ge(1) atom,  two tensors for the other bonds are expressed as

\begin{equation}
\mathbb{A}^\mathrm{Ge(1)}_2 = \frac{1}{4}\left[\begin{array}{ccc}
a_\perp+3a_\parallel & \sqrt{3}(a_\parallel-a_\perp) &  0 \\
\sqrt{3}(a_\parallel-a_\perp) & 3a_\perp+a_\parallel &  0 \\
0 & 0 &  4a_z
\end{array}\right],
 \label{Ge1_tensor}
\end{equation} 

\begin{equation}
\mathbb{A}^\mathrm{Ge(1)}_3 = \frac{1}{4}\left[\begin{array}{ccc}
a_\perp+3a_\parallel & -\sqrt{3}(a_\parallel-a_\perp) &  0 \\
-\sqrt{3}(a_\parallel-a_\perp) & 3a_\perp+a_\parallel &  0 \\
0 & 0 &  4a_z
\end{array}\right].
 \label{Ge1_tensor}
\vspace{0.2cm}
\end{equation} 

\noindent The local transferred hyperfine field at Ge(1) from all three nn Ce atoms reads 

\begin{equation}
%\vspace{0.05cm}
\vec{H}_\mathrm{loc}^\mathrm{Ge(1)}=\sum_{i=1}^3 \mathbb{A}^\mathrm{Ge(1)}_i \cdot \vec{m_i},
 \label{Ge1_tensor}
 \vspace{0.05cm}
\end{equation} 

\noindent where $\vec{m_i}$ is the moment at the $i$-th Ce site. 
For paramagnetic and ferromagnetic states $\vec{m_i}$ is uniform: $\vec{m_i} = \vec{m}$. 
Defining $\vec{m}$ in a polar coordinate, $M(\cos\theta\sin\phi,\ \sin\theta\sin\phi,\ \cos\phi)$, with the moment size $M$ and the polar ($\theta$) and azimuthal ($\phi$) angles, $\vec{H}_\mathrm{loc}^\mathrm{Ge(1)}$ can be written as

\begin{equation}
	 \left(\begin{array}{c}
H_x^\mathrm{Ge(1)}  \\
H_y^\mathrm{Ge(1)} \\
H_z^\mathrm{Ge(1)}
\end{array}\right) = \mathbb{A}^\mathrm{Ge(1)} \cdot M \left(\begin{array}{c}
\cos\theta\sin\phi  \\
\sin\theta\sin\phi \\
\cos\phi 
\end{array}\right),
 \label{Ge1_tensor}
\vspace{0.1cm}
\end{equation}

\noindent with the total nn coupling tensor that is diagonal and axially symmetric:

\begin{equation}
   % \vspace{0.1cm}
    \mathbb{A}^\mathrm{Ge(1)} = (A_{xx}, A_{yy}, A_{zz}) = \left[\begin{array}{ccc}
\frac{3}{2}(a_\perp+a_\parallel) & 0 &  0 \\
0 & \frac{3}{2}(a_\perp+a_\parallel) &  0 \\
0 & 0 &  3a_z
\end{array}\right].
 \label{Ge1_tensor}
\end{equation}

\noindent In the paramagnetic state, when the external field is applied within the $ab$-plane as in our self-aligned powder sample, $A_{xx}$ and $A_{yy}$ can be determined from the slope of the $K$--$\chi$ plot (inset of Fig.~7 in the main text).
In the ferromagnetic state, Eq.~\ref{Ge1_tensor} guarantees that the internal local field at Ge(1) is \textit{exactly} parallel to the direction of the ordered moment ($\vec{H}_\mathrm{loc}^\mathrm{Ge(1)} \parallel \vec{m}$). 
This last point remains valid even if we consider the contributions from other Ce atoms at a longer distance as a direct consequence of the three-fold rotational symmetry at Ge(1) and the mirror plane existing in the hexagonal basal plane.

For the Ge(2) site, similar arguments can be made, yet due to the lower local-site symmetry, ill-defined off-diagonal elements remain that may prevent us from fully determining the relation between $\vec{H}_\mathrm{loc}^\mathrm{Ge(2)}$ and $\vec{m}$. Nonetheless, as the array of Ge(2) atoms lies on a mirror plane [Fig.~1(a) in the main text], the local field $\vec{H}_\mathrm{loc}^\mathrm{Ge(2)}$ necessarily lies within the $ab$-plane when the ordered moment $\vec{m}$ has an easy-plane anisotropy.

\begin{figure}[b]
	\includegraphics[width=7.0cm]{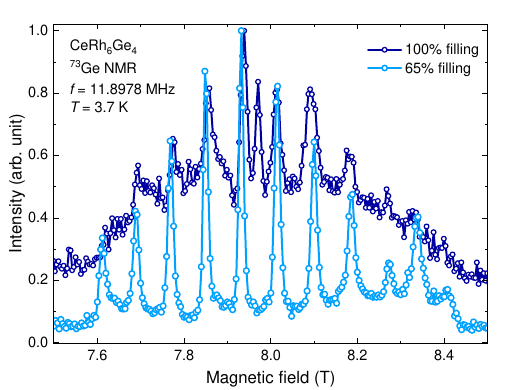}
	%\vskip -0.25cm
	\caption{Field-swept $^{73}$Ge NMR spectrum using powder sample of CeRh$_6$Ge$_4$, recorded at 3.7\,K. The broader spectrum is for the samples packed fully into the NMR coil (100\,\% filled; dark blue), whereas the sharp peak structure is seen for loosely packed samples (65\,\% filled; light blue) that are in-situ aligned ($H$\,$\parallel$\,$ab$) with the external field.}  
	\label{fig:FA}
\end{figure}

\begin{figure}[b]
	\includegraphics[width=8.6cm]{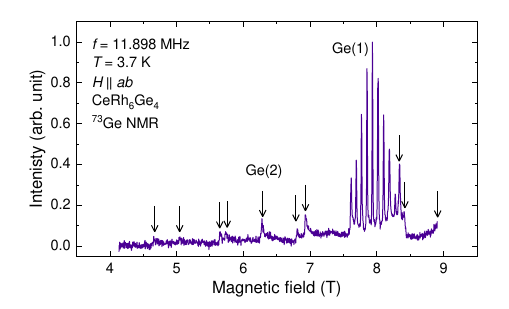}
	\vskip -0.25cm
	\caption{Field-swept $^{73}$Ge NMR spectrum using self-aligned samples in a loosely packed coil (50\,\% filled), recorded at 3.7\,K. The sharp peaks around 8.0~T are from Ge(1), whereas the arrows indicate spectral signatures from Ge(2).} 
\vspace{0.1cm}
	\label{fig:NMR}
\end{figure}

    \subsection{In-situ self-alignment of powder sample and NMR spectrum at Ge(1) for \textbf{\textit{H }$\parallel$ \textit{ab}}}

Figure~\ref{fig:FA} shows a comparison of field-swept NMR spectrum at 3.7\,K around 8.0\,T using an NMR coil that is packed with powder sample with a different filling factor. 
The blue curve corresponds to a closely packed condition with an approximately 100\,\% filling, while the pink curve stands for a 65\,\% filling. 
The curve with a higher filling shows a broader background signal due to both Ge(1) and Ge(2), and a less intense yet more complicated line structure reflecting the random orientation of the crystallites.  
By contrast, the loosely filled curve gives a well-resolved quadrupole-split spectrum with nine NMR peaks from Ge(1) [i.e., eight satellite transitions ($I_z =-9/2 \leftrightarrow -7/2$, $-7/2 \leftrightarrow -5/2$, $-5/2 \leftrightarrow -3/2$, $-3/2 \leftrightarrow -1/2$, $1/2 \leftrightarrow 3/2$, $3/2 \leftrightarrow 5/2$, $5/2 \leftrightarrow 7/2$, and $7/2 \leftrightarrow 9/2$) 
and one central line ($I_z = -1/2 \leftrightarrow 1/2$)], reflecting an in-situ self-alignment of the crystallites due to the easy-plane magnetic anisotropy. 
The small asymmetric tail-like patterns seen at each satellite (towards the central line direction) reflects the powder pattern due to a small fraction of misaligned crystallites. A peak at 8.33\,T and a shoulder around 8.40\,T are signals from Ge(2).
In Fig.~\ref{fig:NMR} we show a field-swept spectrum using self-aligned powder loosely packed into the coil (50\,\% filled), recorded at 4.0\,K over a wider field range from 4.0 to 9.0\,T.
Outside the range where the nine peaks from Ge(1) are visible (between 7.6 and 8.4\,T), several additional peaks and kinks (indicated by arrows) and a broad background are found.  
These features of the spectrum are the powder pattern of Ge(2) due to its finite asymmetry parameter ($\eta^{\rm{Ge(2)}} \neq 0$) and the random orientation of $V^{\rm{Ge(2)}}_{zz}$ ($\parallel ab$) with respect to the external field direction. 
The higher-field half of the full spectrum is missing because of the field limitation of our magnet.

\begin{figure}[!b]
	\includegraphics[width=7.2cm]{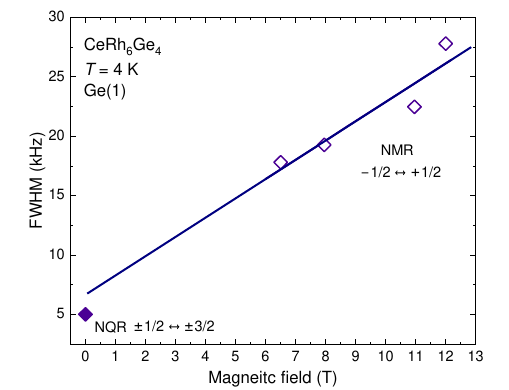}
	%\vskip -0.25cm
	\caption{Full width at half maximum of the central line ($I_z = - 1/2 \leftrightarrow  1/2$) of the $^{73}$Ge NMR spectrum at Ge(1) plotted against external magnetic field (open symbols), recorded at 4.0 K. Field-aligned powder samples ($H$\,$\parallel$\,$ab$) are used. The solid symbol gives the corresponding linewidth of the 1$\nu_\mathrm{Q}$ line ($I_z = \pm 1/2 \leftrightarrow  \pm 3/2$) of the NQR spectrum at Ge(1). Line is a guide to the eyes. }
	\label{fig:HvsFWHM}
\end{figure}

Figure~\ref{fig:HvsFWHM} shows the NMR linewidth's field dependence for the self-aligned powder sample for the central transition line ($I_z = -1/2\leftrightarrow+1/2$) at Ge(1), recorded 4.0\,K. 
In general, the NMR linewidth for quadrupolar nuclei ($I \geq 1/2$) is caused by two factors: 
one is related to magnetic broadening and the other is due to quadrupolar broadening. 
The former reflects the distribution of magnetic field (e.g., a distribution of spin shift), whereas the latter is due to a variation of the electric field gradient at the nuclear position over the sample.
A monotonic increase is visible in the width (Fig.~\ref{fig:HvsFWHM}), which suggests that it reflects a distribution of the shift that scales linearly with the external field. 
The intercept of a linear fit to the data gives $\approx$\,7\,kHz, which is very close to the zero-field NQR linewidth for the $\pm1/2\leftrightarrow\pm3/2$ transition at Ge(1) ($\approx$\,5\,kHz).
The similarity of these two values suggests a high efficiency of our field alignment and reveals that the dominant source of the broadening in the NMR spectrum at higher field and low temperature is the distribution of the shift.

\begin{figure}[!t]
	\includegraphics[width=8.0cm]{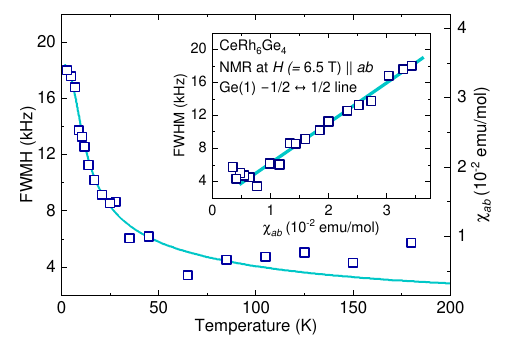}
	%\vskip -0.25cm
	\caption{Temperature dependence of full width at half maximum of the central line ($I_z = - 1/2 \leftrightarrow  1/2$) of the field-aligned powder $^{73}$Ge NMR spectrum of Ge(1) at 6.5~T ($H$\,$\parallel$\,$ab$). 
The solid curve is the corresponding in-plane magnetic susceptibility $\chi_{ab} = M_{ab}/H$ at the same field. Inset shows the width plotted against $\chi_{ab}$.}
\vspace{0.2cm}
	\label{fig:FWHM_NMR}
\end{figure}

Figure \ref{fig:FWHM_NMR} shows the NMR linewidth of the central line ($I_z = -1/2\leftrightarrow+1/2$) of Ge(1) plotted against temperature at 6.5\,T.
Above 100\,K, the linewidth is almost identical to the NQR linewidth of the $\pm1/2\leftrightarrow\pm3/2$ transition, pointing to a small susceptibility at high temperature.
With cooling, the linewidth grows monotonically and increases by a factor of three at 4.0\,K. 
The inset shows the relation between the width and the in-plane magnetic susceptibility measured in another sample from the same batch. 
Linear scaling is observed between the two quantities, indicating that the width is induced by the paramagnetism of Ce 4$f$-electron moment.
The width does not show an anomaly around $T^*$ where we saw a kink in the $K$--$\chi$ plot (Fig.~7 in the main text). 
This suggests that the shift anomaly is not related to inhomogeneity that could otherwise have caused an anomaly in the linewidth--$\chi$ plot, as observed in another Ce-based antiferromagnetic system CeLi$_3$Bi$_2$ that has a similar trigonal structure and an easy-plane magnetic anisotropy~\cite{Bordelon2025}.

\subsection{Single-crystal mosaic NMR spectrum measurements at Ge(2) for  \textbf{\textit{H }$\parallel$ \textit{c}}}

To check the directions of the principal axes of the EFG tensor at Ge(2), we made a mosaic of small single crystals and performed NMR experiments by aligning them along their crystalline $c$-axis direction and applying an external magnetic field along that direction ($H$\,$\parallel$\,$c$).  
For this orientation, three Ge(2) atoms in the unit cell (see Fig.~1 in the main text or the inset of Fig.~\ref{fig:DF}) become magnetically equivalent, which results in nine quadrupole-split NMR peaks.  
Figure~4 in the main text presents the resulting field-swept NMR spectrum recorded at $f=$\,9.70\,MHz and $T=$\,5.0 K for $H$\,$\parallel$\,$c$. 
We can clearly see seven out of nine peaks of the Ge(2) spectrum as expected in a single crystal for $H$\,$\parallel$\,$c$, which indicates that the crystallites are nicely aligned along the $c$-axis direction.
Also shown is a simulation based on exact diagonalization of Eq.~(1) in the main text using the EFG parameters $\nu_{\mathrm{Q}}^\mathrm{Ge(2)} = 1.615$\,MHz and $\eta^\mathrm{Ge(2)} = 0.365$ (the gray pattern in Fig.~4 in the main text). 
The simulation allows us to assign the seven peaks to one central line and six satellites of the Ge(2) spectrum. 
Furthermore, it determines that $V_{zz}^\mathrm{Ge(2)}$ and $V_{yy}^\mathrm{Ge(2)}$ lie within the $ab$-plane while $V_{xx}^\mathrm{Ge(2)} \parallel c$. 
Two Ge(2) satellite peaks are missing due to the field range and signal-to-noise ratio. The Ge(1) spectrum is not visible likely due to the low signal-to-noise ratio and different required pulse conditions.

%\begin{figure}[!h]
	%\includegraphics[width=9.6cm]{FigS11.pdf}
	%\vskip -0.25cm
	%\caption{Field-swept $^{73}$Ge NMR spectrum of Ge(2) (red) using aligned single crystals along the crystalline $c$ axis, recorded at 5.0\,K and $H$\,$\parallel$\,$c$. Shaded pattern (gray) gives the calculated single-crystalline spectrum of Ge(2) using exact diagonalization of Eq.~(1) in the main text and the EFG parameters $\nu_{\mathrm{Q}}^\mathrm{Ge(2)} = 1.615$\,MHz and $\eta^\mathrm{Ge(2)} = 0.365$. The directions of the principal axes of the EFG tensor are assumed to be $V_{zz}^\mathrm{Ge(2)}$ and $V_{yy}^\mathrm{Ge(2)}$ within the $ab$-plane and $V_{xx}^\mathrm{Ge(2)} \parallel c$. The $I_z = -9/2  \leftrightarrow -7/2$ line and the Ge(1) signal are missing due to the low signal-to-noise ratio and field range.} 
	%\label{fig:Ge2-NMR-out-of-plane-field}
%\end{figure}

\subsection{Dipolar field simulation at Ge(1) and Ge(2)}
\begin{figure}[!h]
	\vspace{0.3cm}
    \includegraphics[width=8.0cm]{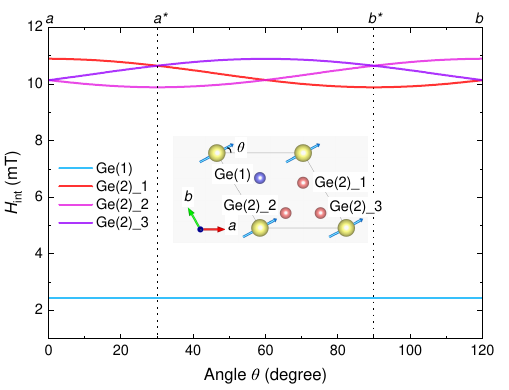}
	%\vskip -0.25cm
	\caption{Calculated direct dipole magnetic field at Ge(1) and Ge(2) from the ordered Ce 4$f$-moments. We assumed uniformly ordered moments of a size 0.3\,$\mu_B$ that are confined within the $ab$-plane. The calculation is done by varying  the angle $\theta$ between the direction of the moments and the $a$-axis within the basal plane and considering Ce atoms lying within a sphere of a radius of 100 \AA.  Inset: Schematic illustration of the crystal structure viewed from the $c$ axis, with the ordered moment shown by arrows. Yellow, purple, and red balls represent Ce, Ge(1), and Ge(2) atoms, respectively, with the magnetically non-equivalent three Ge(2) atoms indicated as Ge(2)\_1, Ge(2)\_2, and Ge(2)\_3.} 
	\label{fig:DF}
\end{figure}

Direct dipole magnetic fields at Ge atoms from the surrounding Ce 4$f$-electron moments are calculated assuming an ordered moment of $m = 0.3\,\mu_B$~\cite{Shen,Matsuoka} lying within a sphere of a radius of 100 \AA~and freely rotating within the $ab$-plane.  
Figure \ref{fig:DF} shows the calculated dipole fields at Ge(1) and Ge(2) plotted as function of the angle $\theta$ defined between the direction of Ce moments and the $a$-axis within the basal plane.
The size of the field at Ge(1) is $\approx$\,2.5 mT and doesn't show any angular variation relative to the moment direction due to the three-fold rotational symmetry. 
The  magnitude of the field varies at Ge(2) with changing $\theta$ due to a lower local-site symmetry and results in maximally three different values ranging between 10 to 11\,mT. 
These values of dipole fields are at least 10 and 5 times smaller than experimentally observed local internal fields at Ge(1) and Ge(2), respectively, as highlighted in Fig.~6(c) in the main text. 
Therefore, we conclude that the local field at the Ge sites is not dominated by dipole fields but predominantly by transferred hyperfine fields.

\newpage